\documentclass[12pt,preprint]{aastex}
\usepackage{emulateapj5}
\usepackage{onecolfloat}

\shorttitle{SIMULATIONS OF DISK GALAXY FORMATION}
\shortauthors{ABADI ET AL.}

\newcommand{\gsim}{ \lower .75ex \hbox{$\sim$} \llap{\raise .27ex \hbox{$>$}} }
\newcommand{\lsim}{ \lower .75ex \hbox{$\sim$} \llap{\raise .27ex \hbox{$<$}} }

\begin{document}

\twocolumn[

\title{ Simulations of Galaxy Formation in a $\Lambda$CDM Universe I: \\
Dynamical and Photometric Properties of a Simulated Disk Galaxy}

\author{Mario G. Abadi\altaffilmark{1} and Julio F. Navarro\altaffilmark{2}}
\affil{Department of Physics and Astronomy, University of Victoria, Victoria, BC V8P 1A1, Canada}

\author{Matthias Steinmetz\altaffilmark{3}}
\affil{Steward Observatory, 933 North Cherry Avenue, Tucson, AZ 85721, USA, and 
Astrophysikalisches Institut Potsdam, An der Sternwarte 16, D-14482 Potsdam, Germany }

\and

\author{Vincent R. Eke\altaffilmark{4}}
\affil{Physics Department, Durham University, South Road, Durham DH1 3LE, England}

\begin{abstract}
We present a detailed analysis of the dynamical and photometric
properties of a disk galaxy simulated with unprecedented numerical
resolution in the $\Lambda$CDM cosmogony. The galaxy is assembled
through a number of high-redshift mergers followed by a period of
quiescent accretion after $z\sim 1$ which lead to the formation of two
distinct dynamical components: a spheroid of mostly old stars and a
rotationally-supported disk of younger stars. The surface brightness
profile is very well approximated by the superposition of an $R^{1/4}$
spheroid and an exponential disk. Each photometric component
contributes a similar fraction of the total luminosity of the system,
although less than a quarter of the stars form after the last merger
episode at $z\sim 1$.  In the optical bands the surface brightness
profile is remarkably similar to that of Sab galaxy UGC615, but the
simulated galaxy rotates significantly faster and has a declining
rotation curve dominated by the spheroid near the center. The decline
in circular velocity is at odds with observation and results from the
high concentration of the dark matter and baryonic components, as well
as from the relatively high mass-to-light ratio of the stars in the
simulation. The simulated galaxy lies $\sim 1$ mag off the I-band
Tully-Fisher relation of late-type spirals, but seems to be in
reasonable agreement with Tully-Fisher data on S0 galaxies. In
agreement with previous simulation work, the angular momentum of the
luminous component is an order of magnitude lower than that of
late-type spirals of similar rotation speed. This again reflects the
dominance of the slowly-rotating, dense spheroidal component, to which
most discrepancies with observation may be traced. On its own, the
disk component has properties rather similar to those of late-type
spirals: its luminosity, its exponential scalelength, and its colors
are all comparable to those of galaxy disks of similar rotation
speed. This suggests that a different form of feedback than adopted
here is required to inhibit the efficient collapse and cooling of gas
at high redshift that leads to the formation of the
spheroid. Reconciling---without fine tuning---the properties of disk
galaxies with the early collapse and high merging rates characteristic
of hierarchical scenarios such as $\Lambda$CDM remains a challenging,
yet so far elusive, proposition.
\end{abstract}

\keywords{cosmology, dark matter, galaxies: formation, galaxies: structure}
]

\altaffiltext{1}{Observatorio Astron\'omico, Universidad Nacional de C\'ordoba and Consejo Nacional de Investigaciones Cient\'{\i}ficas y T\'ecnicas, CONICET, Argentina; abadi@uvic.ca}
\altaffiltext{2}{Fellow of CIAR and of the Alfred P. Sloan Foundation; jfn@uvic.ca}
\altaffiltext{3}{Packard Fellow and Sloan Fellow; msteinmetz@aip.de}
\altaffiltext{4}{Royal Society University Research Fellow; v.r.eke@durham.ac.uk}

\section{Introduction}
\label{sec:intro}

Ever since the photographic plates of extragalactic nebulae were first
systematically examined in the early 20th century it became clear that---despite
their unquestionable individuality---galaxies share certain regularities in
their appearance that make it compelling to organize them into a few broad
morphological classes. The Hubble sequence (Hubble 1926) summarizes this early
taxonomical attempts and identifies several key morphological features
(spheroid, disk, bars, spiral patterns) whose relative prominence conveys a
wealth of information regarding the formation and evolution of individual galaxy
systems.  These features underlie most morphological classification schemes, and
accounting for their origin, for the statistical distribution of galaxies
amongst classes, as well as for their dependence on intrinsic properties (such
as luminosity and rotation speed) and environmental properties (such as
clustering), has become one of the holy grails of galaxy formation studies.

Within the current paradigm of structure formation there is a well
specified scenario for the occurrence and evolution of such
morphological features and, in particular, for the origin of the
spheroidal and disk components. In the simplest version of this
scenario (perhaps the only one that one might be able to rule out
conclusively) most stars in the universe are envisioned to form as a
result of dissipative settling of gas into centrifugally supported
disk-like structures at the center of massive dark matter halos (White
\& Rees 1978). This idea is motivated by the observation that star
forming activity in the local Universe is dominated by the gradual
transformation of gas into stars in spiral disks such as that of the
Milky Way (Kennicutt 1998a, Gallego et al. 1995, Lilly et al 1998,
Sanders \& Mirabel 1996). Centrifugally supported disks are thus the
natural consequence of smooth dissipative collapse and the location
where most stars are born; stellar spheroids arise as the remnants of
subsequent, mainly dissipationless merger events. Galaxy morphology
can thus fluctuate from disk- to spheroid-dominated, a continually
evolving feature in the lifetime of a galaxy driven by the mode and
timing of its mass accretion history.

A small (but significant) fraction of stars, however, are also being formed at
present in starbursting and peculiar galaxies typified by ongoing mergers, and
there is strong indication that the fraction of stars formed in such events was
significantly higher in the past (Steidel et al. 1999).  This point
notwithstanding, the simple scenario outlined above linking accretion and
morphology has found favor in theoretical and simulation work, starting with
the pioneering work of Larson and the Toomres in the 70s (Larson 1977, Toomre
\& Toomre 1972, Toomre 1977), as well as in more recent work where attempts are
made to account for the cosmological context of the mass accretion history of a
galaxy (Katz \& Gunn 1991, Navarro \& Benz 1991, Katz 1992, Evrard, Summers \&
Davis 1994, Navarro \& White 1994, Steinmetz \& M\"uller 1995, 
Steinmetz \& Navarro 2002). This whole body of work has shown
that, qualitatively at least, the origin of multiple dynamical components in
galaxies can be ascribed to the varied mass accretion history characteristic of
hierarchically clustering universes.

Additional support for this scenario comes from semianalytic studies of galaxy
formation which, using plausible (if perhaps non-unique) rules for assigning
morphological types to remnants of merger events, are able to reproduce the
abundance of galaxies as a function of spheroid-to-disk ratio as well as its
general environmental dependence (Cole et al. 2000, Kauffmann, White \& Guideroni 1993, Somerville \&
Primack 1999).  The success of such accounting, however, relies on a number of free
parameterizations of the poorly-understood interplay between gravitational
collapse, star formation, and the energetics of stellar evolution. Indeed, it is
widely accepted that there is enough freedom in such modeling so as to allow for
the basic observational trends to be reproduced in most viable hierarchical
cosmogonies. Thus, despite the heavy dependence of mass accretion on cosmology
and the exquisite sensitivity of galaxy morphologies to mass accretion history,
galaxy morphologies remain a poorly discriminating tool between competing
cosmological models.

Despite their shortcomings, these studies have reached consensus on a few
general results which seem essential to understanding observed galaxy properties
within a hierarchically clustering formation scenario. For example, one crucial
ingredient is a tight coupling between the energetics of stellar evolution and
that of the interstellar/intergalactic gas. Without such regulating ``feedback''
mechanism, most baryons in hierarchical models would collapse, cool and turn
into stars in the early stages of the collapse hierarchy (White \& Rees
1978). Coupled with the active merging history typical of the assembly of
galaxy-sized systems, this would lead to the majority of baryons in the universe
being locked up in old, dynamically hot spheroids, at odds with observation
(Schechter \& Dressler 1987, Fukugita, Hogan \& Peebles 1998).

Viable models of galaxy formation (and especially of disk galaxy formation) in a
cold dark matter (CDM) dominated universe thus adopt an efficient heating
mechanism (usually ascribed to the energetics of stellar ejecta) that prevents
most baryons from being turned into stars at early times. If the heating
mechanism is energetic enough to blow some of the baryons out of their dark
halos altogether (Adelberger et al. 2002) it might also keep them in the form of
diffuse intergalactic gas and modify their pace of accretion into galaxies so
that they might not necessarily follow the assembly pattern of the dark matter
component. Detaching the accretion histories of gas and dark matter might
actually be necessary to preserve the angular momentum of the baryonic component
during the various merger stages of the hierarchy, thus reconciling the meager
large-scale torques responsible for a galaxy's net rotation with the observed
spin of spiral galaxies (Navarro \& Benz 1991, Navarro \& White 1994, Navarro \&
Steinmetz 1997, Weil, Eke \& Efstathiou 1998).

The main difficulty with this idea is that the efficiency of feedback is
observed to be rather low in most star-forming regions in the local universe
(Martin 1999, MacLow \& Ferrara 1999), apparently insufficient to reconcile
observational trends with the collapse and merging history of galaxy systems in
CDM universes. The possibility remains, however, that the mode of star
formation, and consequently the efficiency of feedback, might be a strong
function of redshift, so that simple rules inspired by observations of nearby
galaxies might not apply at early times.  It is nonetheless clear that the
ultimate success of CDM-inspired galaxy formation models relies heavily on a
realistic and accurate description of feedback.

This is the first in a series of papers where we use direct numerical simulation
to study the galaxy formation process in a $\Lambda$CDM universe under the
simplest possible set of assumptions regarding star formation and feedback. Our
algorithm includes only two free parameters; one expressing the relation between
local dynamical and star formation timescales and another parameterizing the
efficiency of feedback energy in affecting the bulk kinetic energy of gas in
star forming regions. The two parameters are calibrated by matching empirical
relations applicable to normal star forming galaxies in the local Universe. This
approach provides not only insight into the nature of the problems afflicting
hierarchical models of galaxy formation but also helps to guide attempts to
resolve them through a careful quantitative assessment of their
shortcomings. Such assessment might enable us to decide whether the predicament
of CDM-based models can only be solved by resorting to radical modifications to
the basic tenets of the cold dark matter theory, such as envisioned in theories
where dark matter is warm (Sommer-Larsen \& Dolgov 2001, Bode, Ostriker \& Turok
2001) or self-interacting (Spergel \& Steinhardt 2000), or whether the observed
properties of galaxies are the result of complex astrophysics prevailing over
some of the ``natural'' trends of gravitational instability models.

\begin{figure*}[t]
\plotone{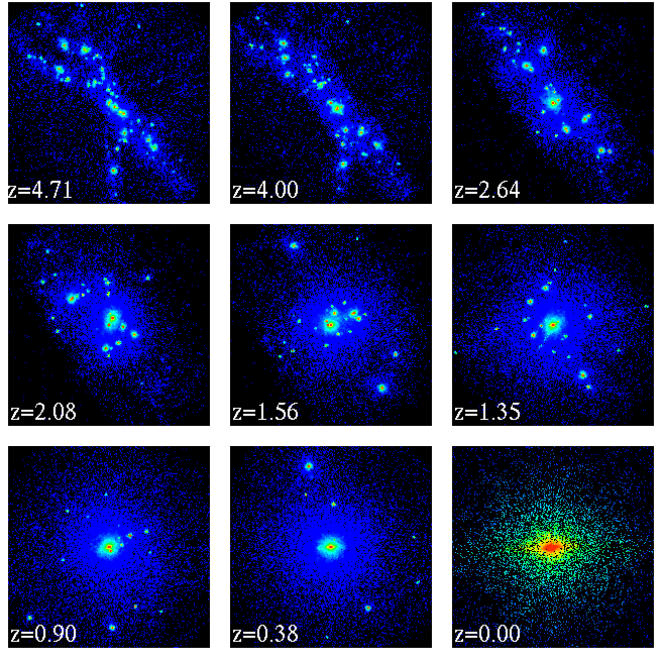}
\caption{
Dark matter particles within a cube of $320$ physical kpc on a side, shown at
various redshifts and projected so that the luminous galaxy at $z=0$ is seen
edge-on. The bottom right panel zooms into the innermost $40$ kpc of the system.
ach particle is colored according to the logarithm of the local dark matter
density. Red and blue correspond to $\rho_{\rm dm} \, \gsim \, 10^{10} \,
M_{\odot}$/kpc$^3$ and $\rho_{\rm dm} \, \lsim \, 10^{6} \, M_{\odot}$/kpc$^3$,
respectively.  }
\label{figs:drk}
\end{figure*}

We present here a detailed analysis of the dynamical and photometric properties
of a disk galaxy simulated in the $\Lambda$CDM cosmogony. The simulation is
similar in many respects to that presented by Steinmetz \& Navarro (2002),
except that in this case the halo has been selected so that it undergoes no
major mergers after $z\sim 1$. This is conducive to the formation of a disk-like
component by gradual accretion of cooled gas in the relatively undisturbed
potential well of the dark halo. As we discuss below, a spheroid of old stars
and a disk of young stars are clearly discernible at $z=0$, supporting the view
that the origin of various dynamical and photometric components in disk galaxies
can be traced directly to events in the mass accretion history of the
galaxy. Although the analysis in this paper deals with a single simulation, we
plan to extend this work to other systems in further papers of this series,
whose final goals include: (i) building up a statistically significant set of
simulations exploring the morphologies of galaxies in halos formed through
various merging histories, (ii) testing the dependence of the
accretion-morphology link on simulation parameters, and (iii) using this
information to predict the abundance of galaxies as a function of morphology in
different cosmological models.

The plan of this paper is as follows. In \S~\ref{sec:numexp} we present details
of the numerical simulation, \S~\ref{sec:results} lists the main results and
compares them with observational data; \S~\ref{sec:disc} discusses the
implications of the modeling for our current understanding of the formation of
disk galaxies in a hierarchical universe, whilst \S~\ref{sec:conc} summarizes
our main conclusions.

\begin{figure*}[t]
\plotone{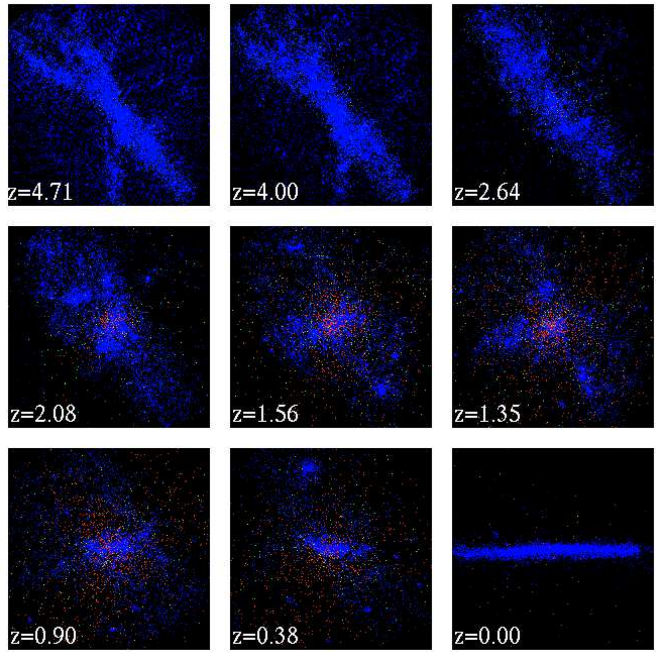}
\caption{
Gas particles within a cube of $320$ physical kpc on a side, shown at different
redshifts and projected so that the luminous galaxy at $z=0$ is seen
edge-on. The bottom right panel zooms into the innermost $40$ kpc of the system.
Each particle is colored according to its temperature. Red and
blue correspond to $T \, \gsim \, 5 \times 10^5 $ K and $T \, \lsim \, 3
\times 10^4 $ K, respectively.}
\label{figs:gas}
\end{figure*}

\section{The Numerical Experiments}
\label{sec:numexp}

\subsection{The Code}
\label{ssec:code}

The simulation described here was performed with GRAPESPH, a particle-based,
fully three-dimensional Lagrangian hydrodynamical code that combines the
flexibility and adaptability of the Smoothed Particle Hydrodynamics technique
with the speed of the special-purpose hardware GRAPE for computing gravitational
interactions (Steinmetz 1996). The version used here includes the self-gravity
of gas, stars, and dark matter, hydrodynamical pressure and shocks, Compton and
radiative cooling, as well as the heating effects of a photoionizing UV
background (see Navarro \& Steinmetz 1997, 2000b for more details).

\subsubsection{Star Formation and Feedback}
\label{sssec:sf}

Star formation is handled in GRAPESPH by means of a simple recipe for
transforming gas particles into stars.  The star formation algorithm is similar
to that described in Steinmetz \& M\"uller (1994, 1995, see also Katz 1992 and
Navarro \& White 1993), where star formation is modeled by creating ``star''
particles in collapsing regions that are locally Jeans-unstable at a rate given
by $\dot{\varrho}_{\star}=c_{\star} \, \varrho_{\rm gas}/\max(\tau_{\rm
cool},\tau_{\rm dyn})$. Here $\varrho_{\rm gas}$ is the gas density and
$\tau_{\rm cool}$ and $\tau_{\rm dyn}$ are the local cooling and dynamical
timescales, respectively. The proportionality parameter, $c_{\star}=0.05$, is
chosen so that in dense regions, where $\tau_{\rm cool}\ll\tau_{\rm dyn}$,
eligible gas is transformed into stars on a timescale much longer than $\tau_{\rm
dyn}$.

After formation, star particles are only affected by gravitational forces, but
they devolve energy and mass to their surroundings, in a crude attempt to mimic
the energetic feedback from evolving stars and supernovae. To be precise, for
the $3 \times 10^7$ yrs following their formation star particles inject into
their surrounding gas $10^{49}$ ergs per solar mass of stars formed. The bulk of
this energy is invested into raising the internal energy (temperature) of the
gas, a rather inefficient way of regulating star formation in dense, cold
environments where cooling timescales are so short that the feedback energy is
almost immediately radiated away.

We attempt to generalize this formulation by assuming that a certain fraction,
$\epsilon_v$, of the feedback energy is invested in raising the kinetic energy
of the neighboring gas. These motions are still dissipated by shocks, but on a
longer timescale, allowing for lower star formation efficiencies and longer
effective timescales for the conversion of gas into stars as $\epsilon_v$
increases.  As discussed by Navarro \& Steinmetz (2000b, hereafter NS00b), we have
attempted to determine plausible values for the two free parameters in our star
formation algorithm, $c_{\star}$ and $\epsilon_v$, by matching the star
formation properties of isolated disk galaxy models with the empirical relation
between star formation rates and gas density compiled by Kennicutt (1998b).

The simulation reported here adopts $c_{\star}=0.033$ and $\epsilon_v=0.05$. As
discussed by NS00b, these choices result in a moderately efficient feedback
scheme, where the choice of $c_{\star}$ prevents the rapid transformation of
cold gas into stars but the low value of $\epsilon_v$ results in only a minor
fraction of cooled gas being reheated and returned to intergalactic space in
diffuse form. As such, we expect the star formation history in our simulated
galaxy to trace roughly the rate at which gas cools and collapses to the center
of dark matter halos.

\begin{figure*}[t]
\plotone{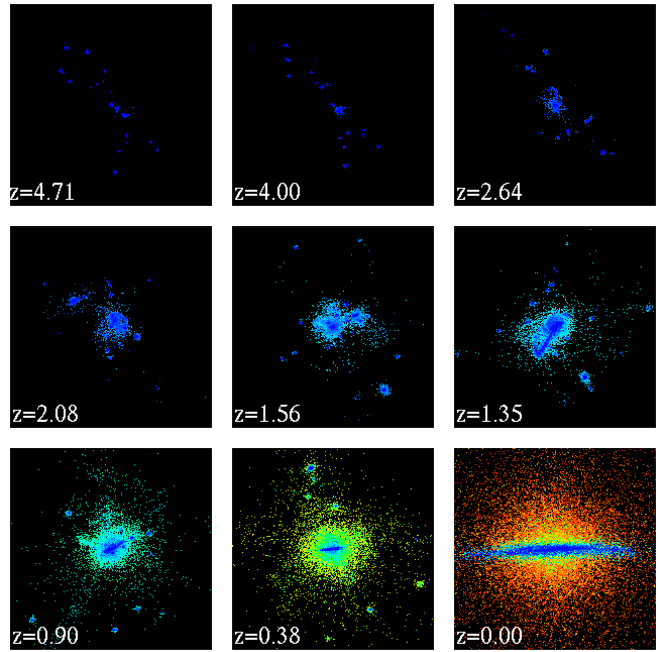}
\caption{ 
Star particles within a cube of $320$ physical kpc on a side, shown at different
redshifts and projected so that the luminous galaxy at $z=0$ is seen
edge-on. The bottom right panel zooms into the innermost $40$ kpc of the system.
Each particle is colored according to its age at the time shown. Blue and red
correspond to $\tau \, \lsim \, 4$ Gyr and $\tau \gsim \, 10$ Gyr, respectively.
}
\label{figs:str}
\end{figure*}

A crude measure of the metal content of gas and stars is tracked by the code
under the assumption that stellar evolutionary processes devolve $1.7\,
M_{\odot}$ of heavy elements to the interstellar medium per $100 \, M_{\odot}$
of stars formed.  This material is added to the gas particles surrounding young
``stars'' for about $3 \times 10^7$ yrs following their formation. The total metal
content of a gas particle is used as a crude measure of its metallicity, which
is in turn inherited by the star particles it spawns. Because of the short
timescale adopted for this process, metallicities in our code are best
understood as reflecting the $\alpha$-element enriched metal pollution due to
type II supernovae rather than the Fe-rich enrichment expected from type Ia
supernovae. No attempt has been made to follow the process of diffusion of
metals in the gas.

It is clear from this discussion that our treatment of metal enrichment is quite
rudimentary and, therefore, we shall restrict the use of stellar metallicities
to the spectrophotometric modeling of the contribution of each star particle to
the luminous output of a galaxy (see \S~\ref{ssec:anal}).  We emphasize that, at
present, the numerical treatment of star formation, feedback, and metal
enrichment is rather uncertain, and there exist a number of implementations
proposed in the literature (see, e.g., Gerritsen 1997, Katz, Weinberg \&
Hernquist 1996, Thacker \& Couchman 2000, Springel 2000, Sommer-Larsen, G\"otz
\& Portinari 2002, Marri \& White 2002). In particular, it is possible that
implementations that include substantial modifications to the kinetic energy of
the gas surrounding star forming regions may lead to rather different star
formation histories than reported here. Thus, the results reported here are best
regarded as preliminary and should be corroborated by experiments designed to
explore other plausible ways of accounting for the transformation of gas into
stars and for their feedback on the interstellar medium.

\subsection{The Simulation}
\label{ssec:sim}

We focus the numerical resources of the simulation on a region that evolves to
form, at $z=0$, a galaxy-sized dark matter halo in the low-density, flat Cold
Dark Matter ($\Lambda$CDM) scenario (Bahcall et al.. 1999). This
is currently the favorite amongst hierarchical clustering models of structure
formation, and is fully specified by the following choice of cosmological
parameters{\footnote{ We express the present value of Hubble's constant as
$H(z=0)=H_0=100\, h$ km s$^{-1}$ Mpc$^{-1}$}}: $\Omega_0=0.3$, $h=0.65$,
$\Omega_{\rm b}=0.019 \, h^{-2}$, and $\Omega_{\Lambda}=0.7$. The power
spectrum is normalized so that at $z=0$ the rms amplitude of mass fluctuations
in $8 \, h^{-1}$\,Mpc spheres is $\sigma_8=0.9$ and we assume that there is no
``tilt'' in the initial power spectrum.

At $z=0$ the dark matter halo under consideration has a circular velocity,
$V_{200}\sim 134$ km/s, and total mass, $M_{200}=5.6 \times 10^{11} \, h^{-1} \,
M_{\odot}$, measured at the virial radius, $r_{200}=134 \, h^{-1}$ kpc, where
the mean inner density contrast (relative to the critical density for closure)
is $200$. This region is identified in a cosmological simulation of a large
periodic box ($32.5 \, h^{-1}$ Mpc on a side) and resimulated at higher
resolution, including the tidal field of the original simulation as described in
detail by Navarro \& White (1994), Navarro \& Steinmetz (1997), and Power et al.
(2002).  The high-resolution region (an ``amoeba''-shaped region contained within
a cube of $3.4 \, h^{-1}$ comoving Mpc on a side) is filled at the initial
redshift, $z_i=50$, with the same number of gas and dark matter particles. The
gas and dark matter particle mass is $m_{\rm g}=2.14\times10^6\, h^{-1}\,
M_{\odot}$ and $m_{\rm dm}=1.21\times 10^7\, h^{-1}\, M_{\odot}$,
respectively. We adopt a Plummer softening scalelength of $0.5$ kpc for all
gravitational interactions between pairs of particles.

The baryonic mass of the final galaxy is roughly $10^{11} M_{\odot}$ (hereafter
we express all $h$-dependent quantities assuming $h=0.65$), equivalent to
roughly $\sim 36,000$ gas particles and represent an order of magnitude
improvement over most previous work.  Few simulations of comparable mass or
spatial resolution have been completed to date, especially considering those
that deal self-consistently with the full three-dimensional hydrodynamical and
gravitational interactions of gas, stars and dark matter within a proper
cosmological context (see, e.g., Thacker \& Couchman 2001, Steinmetz \&
Navarro 2002, Sommer-Larsen, G\"otz and Portinari 2002, Governato et al.
2002). The resolution achieved in this simulation enables a detailed study of
the dynamical and photometric properties of the simulated galaxy, including the
identification of different populations of stars according to age, metallicity,
or kinematics.

\subsection{Analysis}
\label{ssec:anal}

Galaxy luminosities are computed by simply adding up the luminosities of each
star particle, taking into account the time of creation of each particle (its
``age'') and its metallicity, as described in detail by Contardo, Steinmetz \&
Fritze-von Alvensleben (1998). Corrections due to internal absorption and
inclination are neglected, except for a wavelength-dependent dimming intended to
take into account the gradual dispersal of obscuring dust clouds that surround
the formation sites of young stars, and follows closely the prescriptions of
Charlot \& Fall (2000). For example, this amounts to an adjustment of $1.27$ and
$0.93$ mag in the $B$ and $R$ bands, respectively, for stars younger than $10^7$
yrs. For stars older than $10^7$ yrs the adjustment is of $0.42$ and $0.31$ mag
in the same bands, respectively.

\begin{figure}[t]
\plotone{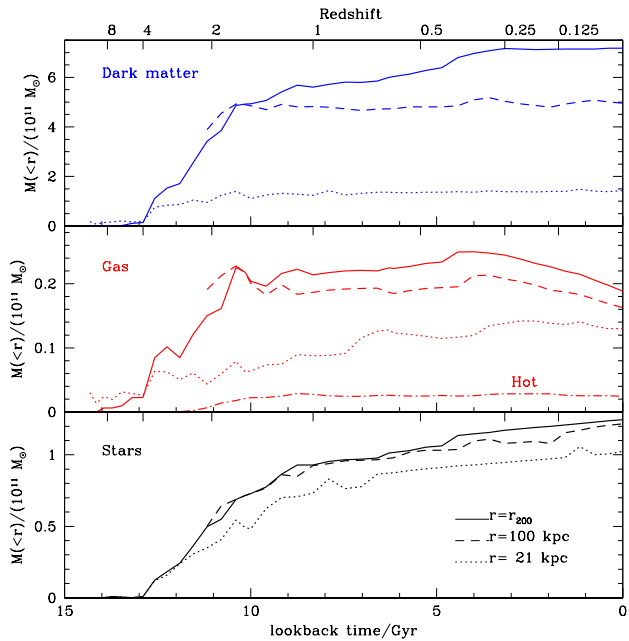}
\caption{Evolution of the mass in the dark matter, gaseous, and stellar
components within various radii. Solid lines indicate the mass within the virial
radius, $r_{200}$, dashed lines correspond to the mass within $100$ physical
kpc, and dotted lines to the mass within $21$ physical kpc, which we define as
the fiducial luminous radius of the galaxy at $z=0$. The curve labelled ``hot'' in
the gas panel corresponds to the total mass in the hot, diffuse component within
$r_{200}$.  }
\label{figs:m200}
\end{figure}

\section{Results}
\label{sec:results}

\subsection{Evolution}
\label{ssec:evol}

Figures~\ref{figs:drk}-~\ref{figs:str} show the distribution of dark matter,
gas, and star particles in the simulation at various times during the
evolution. Each panel is centered on the most massive progenitor and is $320$
kpc (physical) on a side except for the bottom right one, which zooms into the
inner $40$ kpc of the final system at $z=0$. Dark matter particles (Figure 1)
are colored according to density (blue and red correspond to $\rho_{\rm dm} \,
\lsim \, 10^6 \, M_{\odot}$/kpc$^3$ and $\rho_{\rm dm} \, \gsim \, 10^{10} \,
M_{\odot}$/kpc$^3$, respectively), gas (Figure~\ref{figs:gas}) according to
temperature (blue and red correspond to $T \, \lsim \, 3 \times 10^3$ K and $T
\, \gsim \, 5 \times 10^5$ K, respectively), and stars (Figure~\ref{figs:str})
according to their current age (blue and red correspond to $\tau \, \lsim \, 4$
Gyrs and $\tau \, \gsim \, 12$ Gyrs, respectively).

The formation path depicted in Figure~\ref{figs:drk} is fairly typical of the
assembly process of dark matter halos in the $\Lambda$CDM cosmogony. By $z\sim
5$ the high-resolution region collapses into a sheet-like structure crisscrossed
by filaments traced by dark matter halos. Gas is pulled into these non-linear
structures, where it cools, condenses at the center of the non-linear dark
matter clumps, and starts forming stars. Gas in the filaments has short cooling
times, and remains fairly cold throughout; only gas outside of the main
``sheet-like'' structure heats up to roughly a million degrees. The non-linear
clumps are slowly merged together as matter drains down the filamentary
structure into the most massive progenitor.

The merging activity is largely over by $z\sim 1$, when most of the mass is
finally in place. This is shown quantitatively in the top panel of
Figure~\ref{figs:m200}, where the mass of the dark halo measured within various
radii is shown as a function of time. The solid line corresponds to the dark
mass within the virial radius, $r_{200}$, and shows that the mass of the most
massive progenitor has approximately doubled since $z\sim 2$. Most of the
increase happens between $z\sim 2$ and $1$; $M_{200}$ grows by only $30\%$ since
$z=1$. The latter increase is largely a result of the time-dependent definition
of the virial radius. Indeed, the dark mass within the inner $100$ physical kpc
(dashed lines in Figure~\ref{figs:m200}) increases by less than $5\%$ in the
last $10$ Gyr (since $z \sim 1.5$), and that within $r_{\rm lum}=21$ physical
kpc{\footnote{Throughout the paper we adopt $r_{\rm lum}=21$ kpc as a fiducial
``luminous radius'' to define the luminous component of the galaxy at the center
of the dark matter halo.}} (dotted lines) hardly increases at all since then.

This implies that over the past $10$ Gyr the environment where the
luminous component of the galaxy grows is an exceptionally quiet one
conducive to the formation of a disk through the gradual accretion of
cooled gas. The bottom two panels of Figure~\ref{figs:m200} show that
the baryonic mass within $r_{\rm lum}$ increases by about $30\%$ since
$z\sim 1$ as cooled gas gradually settles into a centrifugally
supported disk at the center of the halo. This gas is turned into
stars at about the same rate as it is accreted from the surrounding
halo, so that the fraction of baryonic material in gaseous form in the
central galaxy is always of order $10$-$13\%$. Interestingly, as shown
by the dot-dashed line in Figure~\ref{figs:m200}, little of the gas
that makes up the disk cools off from a quasi-equilibrium corona of
gas in hydrostatic equilibrium, as is commonly assumed in semianalytic
models of galaxy formation (see, e.g., White \& Frenk 1991). Most of
the gas that forms the disk is cold throughout the accretion process
and reaches the center either in precollapsed systems such as
satellites or along dense filaments of mass (see
Figure~\ref{figs:gas}).

\begin{figure}[t]
\plotone{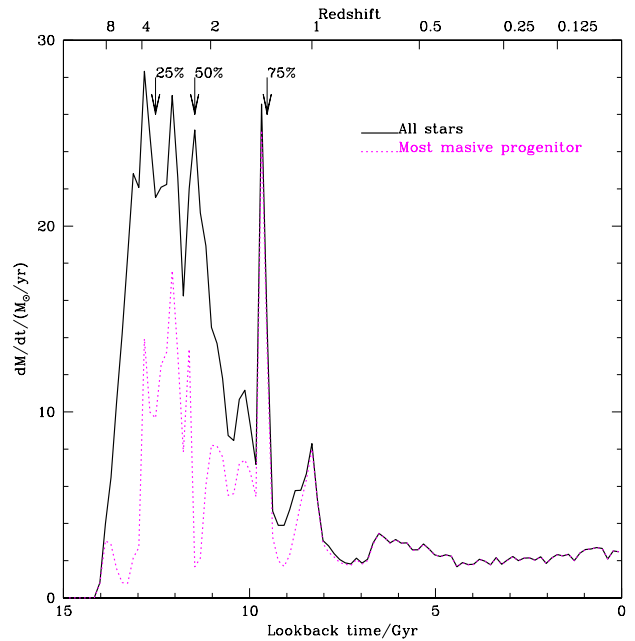}
\caption{The distribution of stellar ages corresponding to stars identified
within $r_{\rm lum}=21$ kpc at $z=0$.  Downward-pointing arrows indicate the
formation times of the first $25\%$, $50\%$, and $75\%$ of stars,
respectively. Peaks in the distribution are easily traced to major merger events
between progenitors of the final galaxy, the last of which happens at $z\sim
1$. Afterwards the star formation rate stabilizes at around $\sim 2\,
M_{\odot}$/yr, which corresponds roughly to the rate of smooth accretion of
cooled gas within the luminous radius of the galaxy.  The dotted line
corresponds to the stars formed within the luminous radius of the most massive
progenitor. About half of the stars formed earlier than $z\sim 1.5$ are formed
in progenitors that are later merged within the final galaxy.}
\label{figs:sfr1}
\end{figure}

The accretion rate of gas onto the center of the halo during the past $\sim
7$-$8$ Gyr is approximately constant and as a consequence the star formation
rate of the galaxy remains at roughly $\sim 2 \, M_{\odot}$/yr over the same
period (Figure~\ref{figs:sfr1}). Most of these stars form and remain in a
centrifugally supported, thin disk-like structure so that, at $z=0$, the central
galaxy has two easily distinguishable components; a spheroid composed mainly of
old stars formed before the merging activity tapers off at $z\sim 1$, and a thin
stellar disk of young stars (seen edge-on in the bottom right panel of
Figure~\ref{figs:str}).

At earlier times the star formation rate is significantly higher, as shown by
the age distribution of all stars identified within $r_{\rm lum}$ at $z=0$
(solid line in Figure~\ref{figs:sfr1}). The dotted curve corresponds to stars
formed exclusively in the central galaxy of the most massive progenitor halo; at
$z\sim 4$ the star formation rate was typically of order $10$-$15 \,
M_{\odot}$/yr, a factor of $\sim 5$ times higher than at present. The age
distribution of all stars also indicates that stars form very efficiently at
early times, reflecting the large amount of gas that is able to cool within the
many progenitors of the final system as well as the weak effect of our feedback
implementation in preventing cooled gas from turning swiftly into stars.

Roughly $25\%$ of the stars in the central galaxy at $z=0$ have
already formed by $z\sim 3.4$, about half of them by $z\sim 2.3$, and
$75\%$ by $z\sim 1.3$, before the last major merger of the galaxy, at
$z\sim 1$. The age distribution is punctuated by clearly discernible
``peaks'' of activity which may be traced to individual merger
events. At $z=0$ the ratio of current to past-average star formation
rate is $\approx 0.3$, consistent with that of spirals of type later
than Sa (Kennicutt 1998a). The similarity between the simulated galaxy
and early type spirals extends beyond its star formation history to
its photometric properties, as we discuss next.

\begin{figure}[t]
\plotone{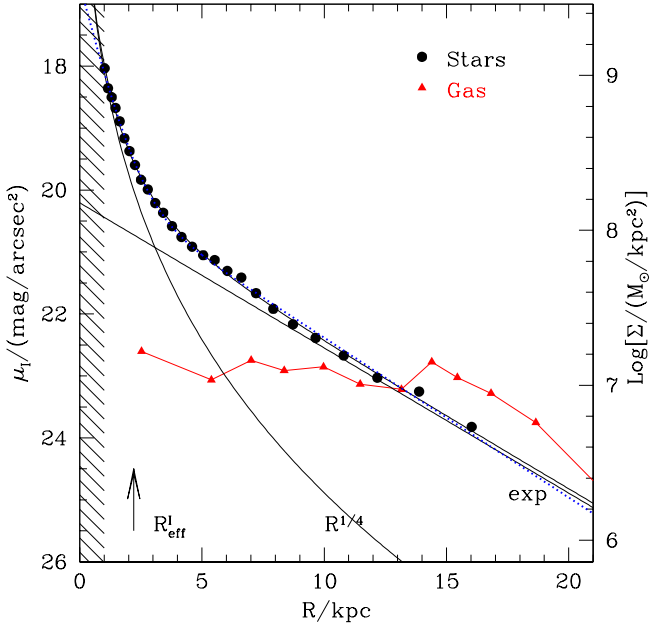}
\caption{
$I$-band surface brightness profile of the galaxy seen projected face-on. Filled
circles correspond to the profile measured for the simulated galaxy; the solid
lines indicate the decomposition into an $R^{1/4}$ spheroidal component and an
exponential disk, as labelled in the figure. The dotted line through the filled
circles correspond to a fit assuming that both the spheroid and the disk have
exponential profiles. The solid triangles correspond to the gas surface density
profile (scale at right). The upward-pointing arrow indicates the projected
half-light radius of the galaxy. The shaded area highlights the region
compromised by our spatial resolution.}
\label{figs:surf1}
\end{figure}

\subsection {Photometric properties}
\label{ssec:photom}

The solid circles in Figure~\ref{figs:surf1} show the $I$-band surface
brightness profile of the simulated galaxy seen face-on. The shape of the
profile is reminiscent of that of spiral galaxies (see, e.g., Boroson 1981,
MacArthur, Courteau \& Holtzman 2002); roughly exponential in the outer parts and with a sharp
upturn towards the center due to the spheroid. The solid line through the filled
circles is the result of a fit that combines a de Vaucouleurs $R^{1/4}$ law and
an exponential profile. This four-parameter fit provides a very good description
of the stellar luminosity profile over $6$ magnitudes, from the inner kpc out to
$\sim 15$ kpc. We emphasize that the success of the $R^{1/4}+$exponential fit
should not be taken to imply that the disk component is necessarily exponential
or that the spheroidal component is necessarily $R^{1/4}$; indeed, assuming
exponential profiles for both the spheroid and the disk results in a fit of
comparable quality (see dotted line in Figure~\ref{figs:surf1}). We shall
explore the correspondence between dynamical and photometric components in
detail in the next paper of this series.

The parameters of the two-component fits for different bands is given in
Table~\ref{tab:gxphotdec}, and indicate that approximately half of the total
light is assigned through this procedure to each component.  The (extrapolated)
central surface brightness of the disk is $22.1$ mag/arcsec$^2$ in the $B$ band,
not very different from the canonical Freeman value of $21.7$ mag/arcsec$^2$
(see, e.g., Freeman 1970, Binney \& Merrifield 1998).  The ratio of spheroidal
and disk scalelengths varies from $R_{\rm eff}/R_{\rm d}=0.07$ in the
ultraviolet to $0.23$ in $K$. The values in the redder bands are in in excellent
agreement with observation, although those in the UV and blue bands seem
systematically low (MacArthur, Courteau \& Holtzman 2002). We note that internal
absorption and orientation effects (largely neglected in our analysis) play a
significant role in the determination of these quantities, and may be
responsible for the poorer agreement between simulation and observation in the
UV.

\begin{figure}[t]
\plotone{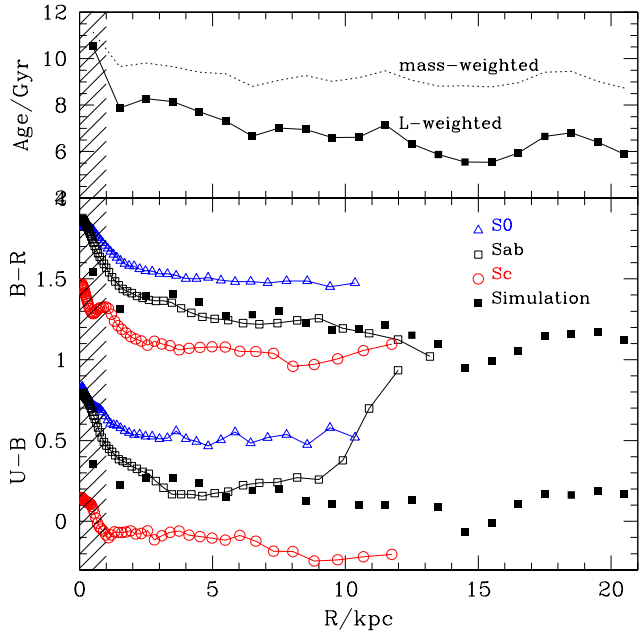}
\caption{Top panel: Age profile of the stellar component of the simulated galaxy
seen face on. Solid squares indicate the luminosity-weighted age in circular
radial bins; the dotted line corresponds to mass-weighted ages. Bottom panel:
Color profiles of the simulated galaxy (filled squares) compared with Sab galaxy
UGC615 (open squares), S0 galaxy UGC10097 (open triangles), and Sc galaxy
UGC10956 (open circles). The colors of the simulated galaxy are too red to be
consistent with an Sc and too blue to be consistent with an S0. There is also a
mild color gradient towards bluer colors in the outer regions consistent with
the age gradient shown in the top panel. The shaded area highlights the region
compromised by our spatial resolution. }
\label{figs:colorsage1}
\end{figure}

The effective radius of the fitted spheroidal component is quite small ($\lsim
\, 1$ kpc) and increases systematically towards longer wavelengths, from $0.4$
kpc in the ultraviolet to about $1$ kpc in the near infrared. The exponential
scalelength of the disk component shows the opposite behaviour, decreasing from
$5.5$ kpc in the $U$ band to about $4.4$ kpc in $K$. These systematic trends
with wavelength are broadly consistent with observation and may be traced to
radial gradients in the average stellar age (see top panel of
Figure~\ref{figs:colorsage1}).  The filled squares in this figure indicate the
luminosity-weighted age of stars as a function of radius, and show that the
oldest (reddest) stars dominate the light in the central kpc, whereas younger
stars are responsible for much of the light in the outer regions.

The distribution of gas differs strongly from the stars and is distributed
across the disk with roughly constant surface density, as is commonly found in
spiral galaxies. Within $r_{\rm lum}$, stars are being formed at a rate of $1.4
\times 10^{-3} \, M_{\odot}\,$kpc$^{-2}\,$yr$^{-1}$ which, together with an average gas
surface density of $\sim 1.1 \times 10^7 \, M_{\odot}\,$kpc$^{-2}$, place the
simulated galaxy in close agreement with the empirical ``Schmidt-law'' relation
described by Kennicutt (1998b) for nearby spirals.

\begin{figure}[t]
\plotone{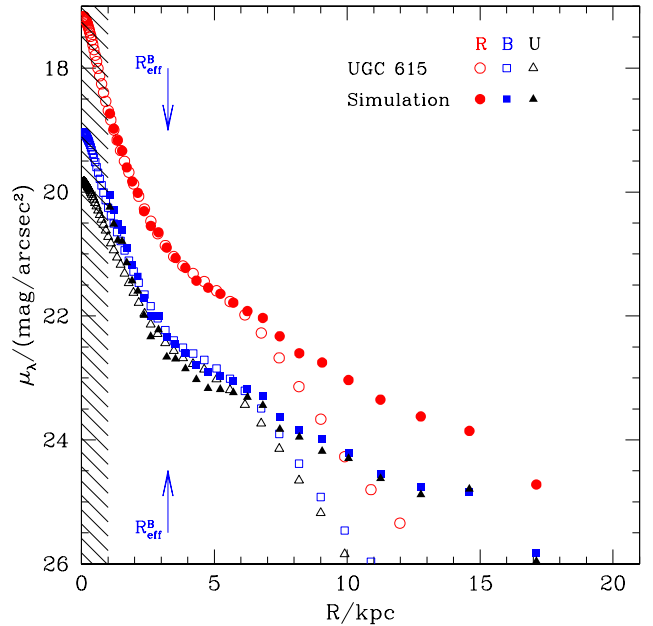}
\caption{Surface brightness profile in the $U$, $B$, and $R$ bands. The
simulated galaxy is represented with filled symbols and is compared with Sab
galaxy UGC615 (open symbols, data from Jansen et al. 2000). Upward- and
downward-pointing arrows indicate the projected half-light radius of the
simulated galaxy and of UGC615, respectively. Note the excellent agreement in
the optical bands between the simulation and UGC615, which extends beyond two
effective radii. The two galaxies have similar luminosities and effective radii,
and {\it there is no rescaling} between observational and simulated data, so the
agreement is genuine. The agreement is poorer in the ultraviolet, possibly due
to internal extinction in UGC615, which is treated crudely in our
modeling. UGC615's light profile is uncharacteristically truncated beyond $\sim
7$ kpc, whereas the exponential decline in the surface brightness of the
simulated galaxy extends out to $\sim 20$ kpc. The shaded area highlights the
region compromised by our spatial resolution.
}
\label{figs:surfubr}
\end{figure}

The similarity between the photometric properties of the simulated galaxy and
those of early-type spirals is further illustrated in Figure~\ref{figs:surfubr},
where we compare the surface brightness profile of the simulated galaxy to that
of UGC615, a nearby Sab galaxy selected from the Nearby Field Galaxy Survey
(Jansen et al. 2000).  We emphasize that {\it there are no arbitrary rescalings}
in this comparison. Near the center, the profiles are remarkably similar,
especially in the $B$ and $R$ bands; in $U$, UGC615's profile is shallower near
the center, again probably as a result of the sensitivity of the UV profile to
internal extinction, which is treated crudely in our analysis. Beyond $\sim 7$
kpc, UGC615's light distribution is (uncharacteristically) truncated; this is
not seen in the simulated galaxy, where the exponential decline in surface
brightness may be traced out to $\sim 20$ kpc.

The color profiles of the simulated galaxy are systematically too blue to be
consistent with S0s and too red compared with Sc's of similar luminosity but
agree very well with UGC615. This is illustrated in
Figure~\ref{figs:colorsage1}, where the color profiles of S0 galaxy UGC10097 and
Sc galaxy UGC10356 are shown with open triangles and circles, respectively. All
of these galaxies exhibit a significant but mild color gradient which is also
nicely reproduced in the simulated galaxy. The radial trend towards bluer colors
in the outer regions is largely due to the age gradient in the simulated galaxy
shown in the top panel of Figure~\ref{figs:colorsage1}. The luminosity-weighted
age drops from $\sim 8$ Gyr at $R=1.5$ kpc (just outside the region affected by
the gravitational softening) to about $\sim 6$ Gyr at $R=15$ kpc. The
mass-weighted age gradient is much less pronounced, as shown by the dotted line
in the top panel of Figure~\ref{figs:colorsage1}.

The radial trend of stellar ages reverses beyond the edge of the star-forming
gaseous disk, and most of the stars that populate the stellar halo (and are not
attached to discernible satellites) are quite old. For example, only $0.9 \%$
($3.4\%$) of stars between $60$ and $150$ kpc are younger than $6$ ($8$)
Gyr. Intriguingly, old stars are also common in the inner 1 kpc (see top panel
of Figure~\ref{figs:colorsage1}): only $3.8 \%$ ($5.5\%$) of stars there are
younger than $6$ ($8$) Gyr. These trends are direct consequences of the
hierarchical assembly of the galaxy, and result from (i) the high-density of the
earliest collapsing progenitors, which makes their central cores resistant to
disruption and allows them to sink to the center of the final remnant; and from
(ii) the large separations of the orbits from which they merge, which allows
tidally-stripped stars to populate orbits that can take them hundreds of kpc
from the center (see Figure~\ref{figs:str}).

We conclude that, overall, the hierarchical scenario of galaxy formation
depicted above is able to reproduce successfully the structural and photometric
properties of early-type spirals such as UGC615. We consider next whether the
agreement extends as well to the detailed kinematical properties of such
galaxies.

\subsection {Dynamical Properties}
\label{ssec:dynprop}

\begin{figure*}
\centerline{\epsscale{2.0}
\plottwo{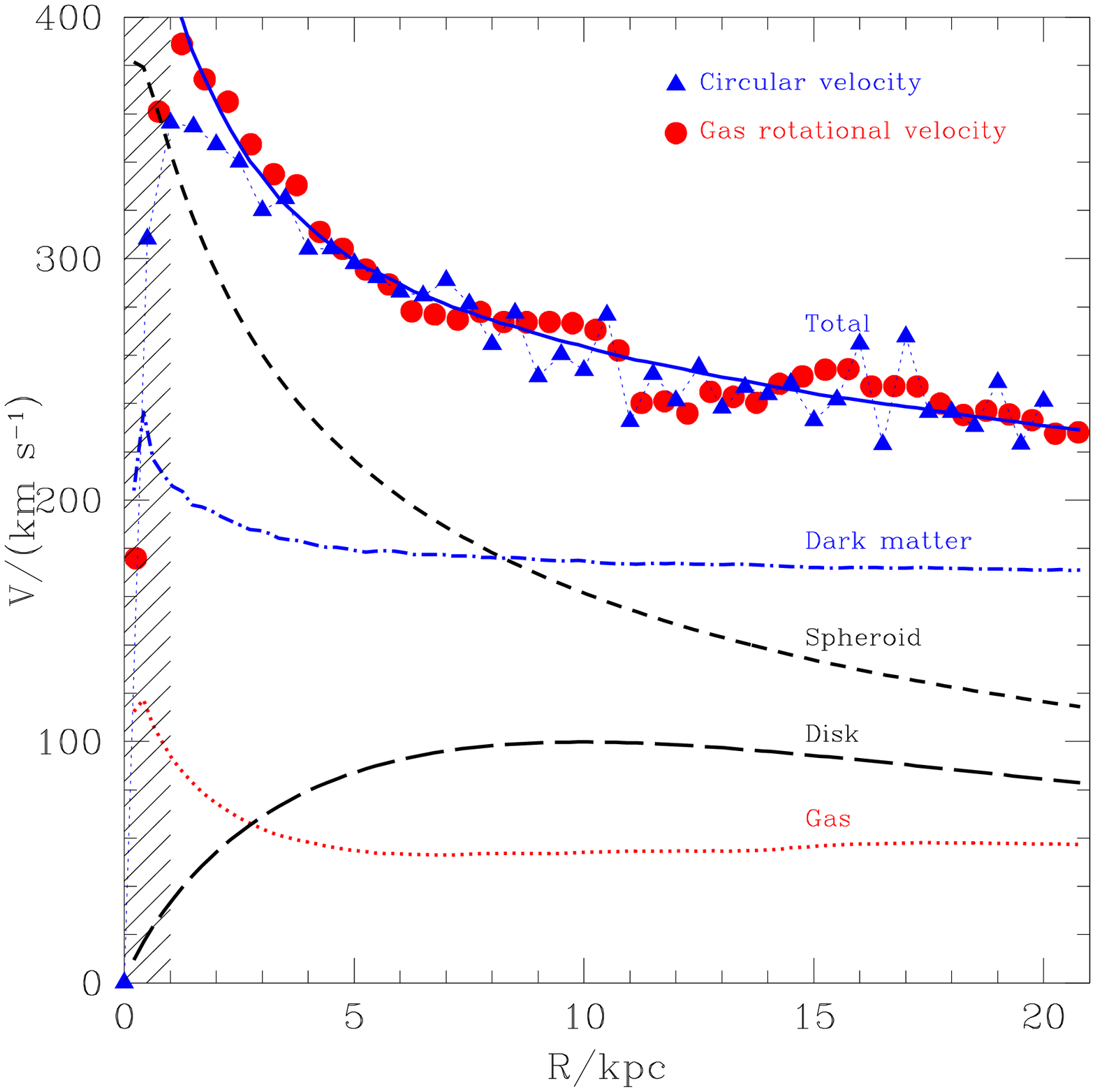}{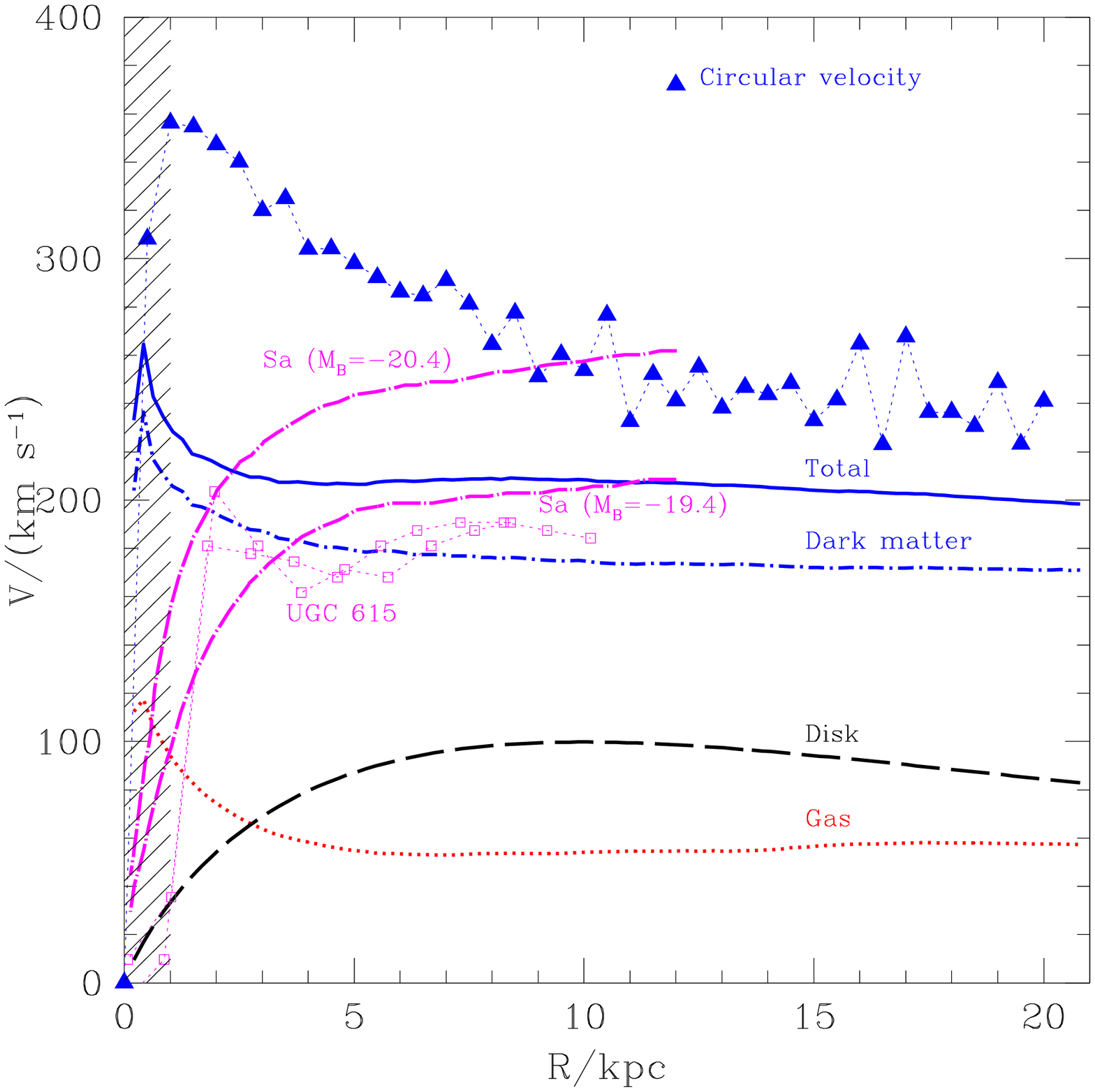}
}
\caption{(a) Left panel. Circular velocity profile of the simulated
galaxy. The (noisy) circular speed curve, estimated by differentiating
numerically the potential on the plane of the disk, is shown with solid
triangles. The gas tangential velocity in the disk follows closely this curve,
and is shown with filled circles. The dark matter constribution to the circular
speed is approximately constant throughout the luminous galaxy, as shown by the
dot-dashed curve. The dashed curves show the contribution of the $R^{1/4}$
spheroid and the exponential disk components derived from the photometric
decomposition shown in Figure~\ref{figs:surf1}. The gas contribution is
indicated by the dotted line. The solid line labelled ``total'' indicates the sum
of the disk+spheroid+gas+dark matter, and is in good agreement with the
true circular speed. (b) Right panel. Same as left panel, but solid line
neglects the contribution of the spheroidal component. The lines labelled ``Sa''
correspond to the rotation curves of Sa galaxies of different luminosities
compiled by Rubin et al. (1985). The (folded) rotation curve of UGC615, kindly
made available by Sheila Kannappan, is shown with open squares. There is poor
agreement between the spheroid-dominated, declining rotation curve of the
simulated galaxy and that of observed galaxies. The shaded area in each panel
highlights the region compromised by our spatial resolution. }
\label{figs:vcir1}
\end{figure*}

The baryonic component of the galaxy dominates the dynamical
properties of the simulated galaxy near the center, mainly as a result
of the high concentration of the luminous component. Half of the total
light from the simulated galaxy comes from within the inner $2.2$ kpc
and $3.3$ kpc in the $I$ and in the $B$ bands, respectively, and is
heavily dominated by the $R^{1/4}$ spheroid near the center. The
luminous component dominates within $\sim 10$ kpc and its spatial
distribution imposes a sharp decline in the circular speed; at $R=1$
kpc the circular speed reaches $\sim 370$ km s$^{-1}$ but it drops to
$260$ km s$^{-1}$ at $R=10$ kpc, as shown in
Figure~\ref{figs:vcir1}. Beyond $R=10$ kpc the circular velocity curve
continues to decline, but at a lower rate, decreasing by only $\sim
12\%$ to $230$ km s$^{-1}$ at $R=20$ kpc. The contribution of the dark
matter to the circular velocity curve is approximately constant across
the galaxy, as shown by the dot-dashed curve in
Figure~\ref{figs:vcir1}.

\begin{figure}[t]
\epsscale{1.0}
\plotone{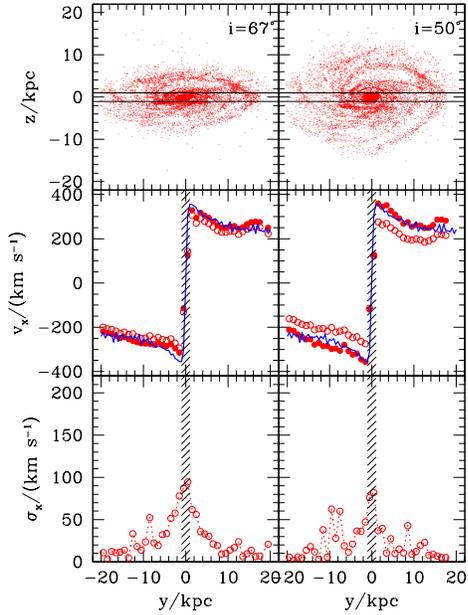}
\caption{
Rotation curve of the gas component of the simulated galaxy, estimated from the
kinematics of gas measured in a $2$ kpc-wide slit aligned with the major axis of
the gas distribution. Panels on the left and right columns correspond to
inclinations of $67$ and $50$ degrees, respectively. The top panels show the
distribution of gas particles. The mean velocity and dispersion across the slit
are shown with open circles in the middle and bottom panels, respectively. The
solid circles in the middle panels correspond to the circular velocity inferred
from these kinematic data after correcting for inclination and asymmetric
drift. After this correction, the inferred circular velocities agree well
with the true circular speed (solid curve), implying that the gas component is
on a thin, centrifugally supported disk with a well behaved velocity field.
}
\label{figs:py_gas}
\end{figure}

The solid circles in Figure~\ref{figs:vcir1} show that the declining
circular velocity curve is closely traced by the gaseous disk.
Indeed, most gas particles are on nearly circular orbits in a thin
disk, as shown in Figure~\ref{figs:py_gas}, where we plot the gas mean
velocity (open circles in middle panel) as well as its dispersion
(bottom panels) measured on a $2$ kpc-wide slit aligned with the major
axis of the galaxy. The left and right panels assume inclinations of
either $67$ or $50$ degrees, respectively. Sixty-seven degrees
corresponds approximately to the highest inclination for which
rotation curves may be derived for thin disks from single-slit
observations without incurring large corrections due to superposition
of different physical radii along the line of sight. The mean measured
velocity along the line of sight is shown by the open circles; the
filled circles show the circular speed, obtained by correcting the
observed velocities by the sine of the inclination angle (derived from
the axis ratio of the gas distribution) and by asymmetric drift,
derived from the velocity dispersion profile using the procedure
outlined by Neistein et al. (1999). The good agreement between the
filled circles and the solid line in Figure~\ref{figs:py_gas} (which
represents the true circular velocity profile, as in
Figure~\ref{figs:vcir1}a) confirms that the gas is distributed on a
thin, cold, centrifugally-supported disk with a very well-behaved
velocity field.

\begin{figure}[t]
\plotone{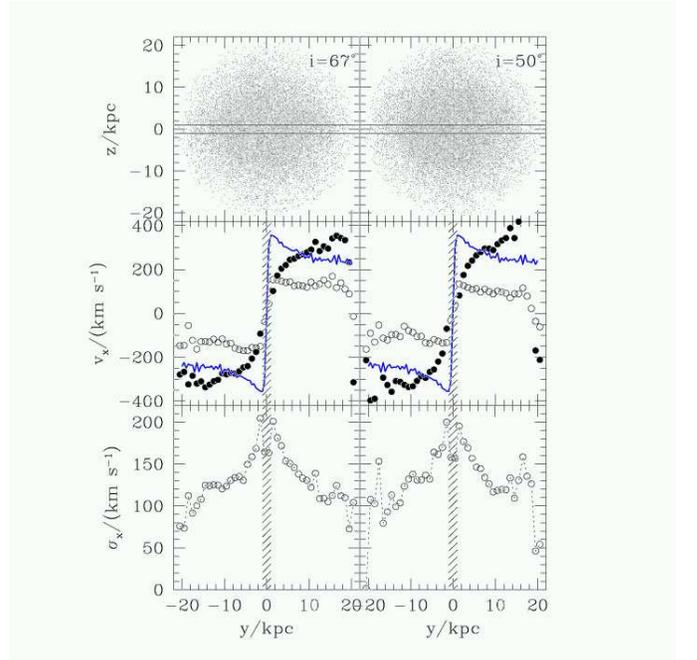}
\caption{ As in Figure~\ref{figs:py_gas}, but for the stellar component. The
velocity dispersion in the stars is comparable to the mean rotation speed at
most radii, making it difficult to estimate circular speeds from single-slit
kinematic information. Indeed, even after correcting for asymmetric drift and
inclination, there is poor agreement between inferred (solid circles) and true
(solid line) circular velocities.}
\label{figs:py_str}
\end{figure}

Young stars (which inherit the dynamical properties of their parent
gas particles)are also on nearly circular orbits in a thin disk, as
seen in the bottom right panel of Figure~\ref{figs:str}. Such stars
contribute a fairly small fraction of the total stellar mass of the
galaxy, which is dominated by the dynamically hot spheroidal
component. This is shown in Figure~\ref{figs:py_str}, where the mean
streaming velocity of the stars is seen to be comparable to the
velocity dispersion at all radii; $V/\sigma$ increases from $\sim 1$
at the center to slightly less than $\sim 1.5$ at $R=20$ kpc from the
center. Such values of $V/\sigma$ are comparable to those in S0
galaxies (Neistein et al. 1999, Mathieu, Merrifield \& Kuijken 2002)
and make it quite challenging to recover true circular speeds from
single-slit spectra of stellar light. Indeed, even after correcting
for asymmetric drift and inclination the recovered circular velocity
curve (shown by the solid circles in Figure~\ref{figs:py_str}) is in
poor agreement with the true circular speed (solid line). This
confirms, as discussed by Neistein et al. (1999), that reliable
information on circular speeds can only be obtained in regions where
rotation dominates so that $V/\sigma$ exceeds at least $\sim
2.5$. This implies that the photometric importance of the disk
component (which contributes half of the total light in this case, see
Table~\ref{tab:gxphotdec}) should be corroborated by dynamical
information seems essential to confirm the prevalence of centrifugally
supported structures in early-type spirals.

The simulated galaxy resembles an early-type spiral in its structural
properties, but its declining circular velocity curve puts it at odds with most
spirals of similar luminosity. For example, as shown in
Figure~\ref{figs:vcir1}b, Sa galaxies in this luminosity range have fairly flat
(rather than declining) rotation curves over the radial range probed by our
simulations (Rubin et al. 1985). In addition, despite the close photometric
similarity between the simulated galaxy and UGC615, the former rotates
significantly faster. The rotation curve of UGC615 (kindly made available by
Sheila Kannappan) is shown with open squares in Figure~\ref{figs:vcir1}b. The
rotation speed peaks at about $190$ km s$^{-1}$, and shows no obvious
decline. Barring some (unlikely) observational artifact that may smear out
rotation speeds in the inner regions, we must conclude that the kinematic
disagreement between the simulated galaxy and UGC615 signals a very different
underlying mass distribution in these two systems.

This is due, at least partially, to the relatively high stellar
mass-to-light ratios that result from our spectrophotometric
modeling. According to the data in Tables~\ref{tab:gxprop} and
~\ref{tab:gxphot}, the stellar mass-to-light ratio of the simulated
galaxy in the $R$ band is $\Upsilon_R=M_{\rm stars}/L_R=2.9 \,
M_{\odot}/L_{\odot}$. This may be compared with the dynamical (i.e.,
{\it maximum}) mass-to-light ratio consistent with UGC615's luminosity
and its rotation speed at the effective radius, $\Upsilon_{\rm max}^e=
G^{-1} V_{\rm rot}^2\, R_{\rm eff}/(C\, L_R)$ where the constant of
proportionality, $C$, depends on the shape of the luminosity profile;
$C=0.42$ for an $R^{1/4}$ spheroid.

For UGC615 we find $\Upsilon_{\rm max}^e=2.05$, about $50\%$ lower than
$\Upsilon_R$ in the simulation. It is thus not surprising that the simulated
galaxy rotates faster than UGC615: both have similar luminosities and half-light
radii but in the former the stars {\it alone} within $R_{\rm eff}$ weigh more
than the maximum mass allowed by dynamical measurements in UGC615.  The
inclusion of the dark matter component, of course, serves only to worsen this
discrepancy, and lifts the velocities in the simulation well above the values
observed for UGC615.  

One might be tempted to vary the IMF in order to lower $\Upsilon$ in
the simulation and bring it into better agreement with the constraints
on UGC615. The values quoted above (and those listed in Tables
~\ref{tab:gxphotdec} and ~\ref{tab:gxphot}) assume a Scalo IMF with
upper and lower mass cutoffs of $100$ and $0.1\, M_{\odot}$,
respectively. For comparison, assuming a Salpeter or Miller-Scalo IMF
would actually make the galaxy {\it dimmer} in $R$ by $\sim 2\%$ and
$\sim 40\%$, respectively. It is not clear to us at this point how to
reconcile the low mass-to-light ratios implied by the dynamical data
of UGC615 with the relatively red colors shown in
Figure~\ref{figs:surfubr} and with the relatively normal ratio of
current to past-average star formation rates (\S~\ref{ssec:evol}). The
disagreement suggests that dynamical constraints such as
$\Upsilon_{\rm max}^e$ should become an important ingredient of
spectrophotometric modeling in order to match observations. We intend
to return to this issue in future papers of this series.

Furthermore, tinkering with the IMF can only make the galaxy brighter
or dimmer as a whole, without resolving the discrepancy in the shape
of the rotation curve. The latter problem is related to the high
concentration of dark matter and baryons in the simulated galaxy
(Navarro \& Steinmetz 2000a). As shown in Figure~\ref{figs:vcir1}, the
contribution of the dark matter to the circular velocity curve is
essentially constant and, therefore, the addition of an $R^{1/4}$
spheroidal component of comparable mass leads inevitably to a
declining rotation curve. Clearly, the dark matter profile in UGC615
and other early-type spirals with flat rotation curves must be
significantly different from that of the simulated galaxy.  We note
that the dark mass distribution responds sensitively to the
concentration of luminous material (Barnes \& White 1984, Blumenthal
et al. 1986), so it is possible that a less massive (or less
concentrated) luminous component would result in a less concentrated
dark matter component in better agreement with observation.

Reconciling simulation and observation thus appears to require lower
stellar mass-to-light ratios than achieved in the simulation but,
perhaps more importantly, also a substantial reduction in the
concentration (or mass) of the luminous galaxy, and of the spheroidal
component in particular. The latter problem is likely related to
mergers of gas-rich progenitors, which are very common at early times
in the history of the simulated galaxy.  As discussed by Mihos \&
Hernquist (1996), gas is funneled very efficiently to the center of
the remnant during mergers, where it may reach extremely high density
before being transformed into stars. Significant modification of our
star formation algorithm appears necessary in order to prevent large
numbers of stars from forming at high redshift in dense configurations
driven by the gas-rich mergers prevalent at that epoch. We elaborate
on this issue further in \S~\ref{sec:disc}.

\subsection{Scaling Laws}
\label{ssec:sclaws}

\subsubsection{Angular momentum}
\label{sssec:angmom}

Accounting for the angular momentum of disk galaxies is one of the
serious challenges faced by hierarchical models of galaxy
formation. The basic problem is that the specific angular momentum of
disk galaxies of given rotation speed is roughly comparable to that of
dark matter halos of similar circular velocity. Since dark matter and
baryons experience similar tidal torquing before turnaround, their
specific angular momentum before collapse is similar, suggesting that
there has been little net transfer of angular momentum between baryons
and dark matter during the assembly of the luminous disk.

On the other hand, numerical simulations where the gas is allowed to radiate and
cool show that baryons tend to collapse early and to cool efficiently within
halos formed during the first stages of the hierarchy (Navarro \& Benz 1991,
Navarro, Frenk \& White 1995, Navarro \& Steinmetz 1997, NS00b). This leads to
significant spatial segregation between components and to substantial transfer
of angular momentum during subsequent mergers from the more centrally
concentrated component (baryons) to the surrounding dark matter. As a result,
the baryonic components of simulated galaxies are found to be deficient in
angular momentum by up to an order of magnitude compared with late-type spirals.

\begin{figure}[t]
\plotone{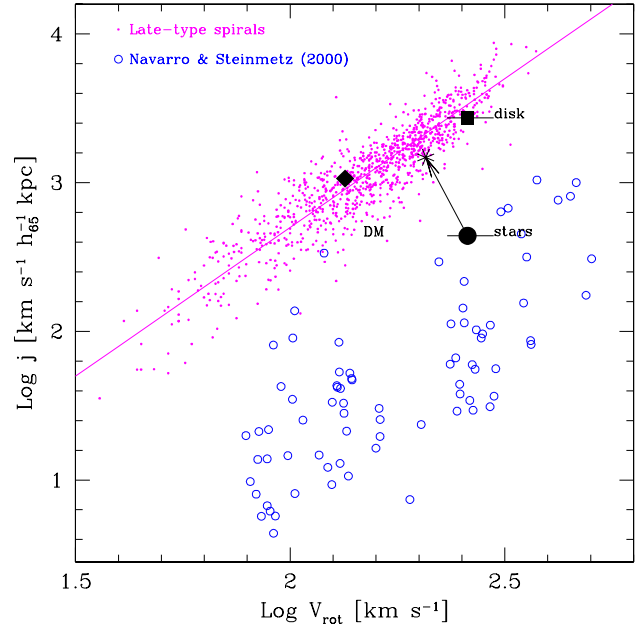}
\caption{Rotation speed versus specific angular momentum. Small filled circles
correspond to late-type spirals taken mostly from the Tully-Fisher samples of
Mathewson, Ford \& Buchhorn (1992) and Courteau (1997), as compiled by Navarro (1998). Open
circles correspond to the simulations of Navarro \& Steinmetz (2000b), which are
similar to the one we present here but of lower resolution. The location of the
dark matter halo in this plane is shown by the filled diamond, assuming $V_{\rm
rot}=V_{200}=134$ km s$^{-1}$. The filled circle correspond to the simulated
galaxy, assuming that the rotation speed is that measured at $R=2.2\, R_{\rm
d}=11$ kpc. The solid square indicates the location of the disk component,
computed adopting the same angular momentum estimator as applied to observed
data, $j=2\,R_{\rm d}\,V_{\rm rot}$. The agreement indicates that the size of
the disk component is similar to that of late-type spirals of similar $V_{\rm
rot}$. The arrow points to where the simulated galaxy would be if the spheroidal
component were removed, as discussed in the text. This confirms that the disk
component of the simulated galaxy is structurally similar to observed late-type
spirals. }
\label{figs:vjdisk}
\end{figure}

A similar problem afflicts the simulation reported here, as shown in
Figure~\ref{figs:vjdisk}. In this figure, the small filled circles correspond to
late-type spirals, taken mainly from the Tully-Fisher samples of Mathewson, Ford \& Buchhorn
(1992) and Courteau (1997), as compiled by Navarro (1998). The solid line is not
a fit to the data but rather the velocity-squared scaling ($j \propto V^2$)
expected for dark matter halos with constant dimensionless rotation parameter
$\lambda$ (see NS00b for details). The agreement in the scaling underlines the
close relationship expected between the angular momentum of baryons and dark
matter. The results of NS00b's numerical simulations are shown with open circles
in Figure~\ref{figs:vjdisk}, and illustrate the angular momentum deficiency of
the baryonic component in simulations alluded to above.

The same is true for the stellar component of the simulated galaxy discussed
here, as shown by the filled solid circle labeled ``stars''. The horizontal ``error
bar'' in this symbol indicates the uncertainty in the rotation speed of the disk
resulting from the declining shape of the rotation curve, which prevents us from
assigning an unambiguous characteristic rotation speed to the simulated
galaxy. We follow standard practice (see, e.g., Courteau 1997) and choose
$V_{\rm rot}=260$ km s$^{-1}$, the velocity at $R=2.2\, R_d=11$ kpc, to
characterize the simulated galaxy. The ``error bar'' spans the range of velocities
in the disk between $R=5$ and $21$ kpc.

The simulated galaxy has significantly lower angular momentum than late type
spirals of comparable rotation speed, due in part to the transfer of angular
momentum from baryons to dark matter during mergers. Indeed, within the virial
radius the dark matter component has twice as much angular momentum as the
baryons, as indicated by the filled diamond labeled ``DM'' in
Figure~\ref{figs:vjdisk}. The disagreement is deepened by the high concentration
of the luminous component, which drives the rotation speed of the disk , $V_{\rm
rot}=260$ km s$^{-1}$, up by a factor of $\sim 2$ over the virial velocity of
the dark matter halo, $V_{200}\sim 135$ km s$^{-1}$.

One important qualification to the angular momentum problem is that specific
angular momenta are quite difficult to measure observationally, and that the
observational data shown in Figure~\ref{figs:vjdisk} assume that the luminous
material is distributed in an exponential disk with a flat rotation curve, for
which $j=2 \, R_d \, V_{\rm rot}$. Computing the angular momentum in the same
manner leads to a much higher estimate of the angular momentum of the simulated
galaxy; for $R_d\approx 5$ kpc and $V_{\rm rot}=260$ km s$^{-1}$ we find
$j\approx 2.6 \times 10^3$ km s$^{-1}$ kpc (for $h=0.65$, see solid square
labeled ``disk'' in Figure~\ref{figs:vjdisk}).  In other words, the {\it size} of
the disk component, estimated from the disk/spheroid photometric decomposition
discussed in \S~\ref{ssec:photom}, is similar to that of late type spirals of
comparable rotation speed. Were the angular momentum of the simulated galaxy not
weighed down by the massive, slowly-rotating spheroidal component, it would
match well that of observed spirals.  This provides further evidence that the
difficulty in reconciling the properties of the simulated galaxy with that of
observed spirals lies in the presence of the dense, slowly-rotating spheroid
that dominates the luminous stellar component.

Finally, we note that the angular momentum of all star particles
formed after the last episode of major merging (i.e., younger than $8$
Gyrs at $z=0$) is $j(\tau < 8$ Gyr$)=1.4 \times 10^3$ km s$^{-1}$ kpc,
about a factor of $\sim 2$ lower than obtained from the $j=2 \, R_d \,
V_{\rm rot}$ estimator. This again highlights the importance of
analyzing observations and simulations using similar techniques, as
well as the crucial role that simulations with numerical resolution
high enough to allow for the proper identification of disk and
spheroidal components will play in resolving the angular momentum
problem of disk galaxies.

\subsubsection{Tully-Fisher relation}
\label{sssec:tf}

\begin{figure}[t]
\plotone{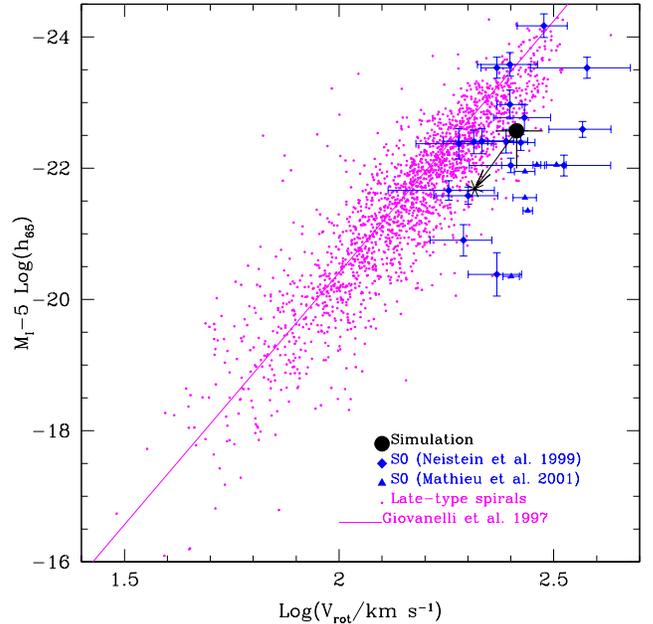}
\caption{ $I$-band Tully-Fisher relation from the samples of Giovanelli et al.
(1997), Han \& Mould (1992), and Mathewson, Ford \& Buchhorn (1992). Diamonds and triangles
correspond to S0s from the work of Neistein et al. (1999) and Mathieu, Merrifield \& Kuijken 
(2002), respectively. The simulated galaxy (filled circle) is seen to lie $\sim
1$ mag fainter than the observational data. This offset is best understood as a
$\sim 20$-$30\%$ velocity offset towards higher speeds caused by the high
concentration of the luminous and dark matter components. The arrow points to
where the simulated galaxy would be if the spheroidal component were removed, as
discussed in the text. The agreement is marginally improved by this exercise,
suggesting that the mass-to-light ratio of the stars might be too high, as
discussed in \S~\ref{ssec:dynprop}.}
\label{figs:tf}
\end{figure}

A powerful diagnostic of the success of simulated galaxies in reproducing
observation comes from comparing their luminosity and characteristic velocity
with the scaling relations linking these properties in galaxies of various
types. The $I$-band Tully-Fisher relation of late type spirals is shown in
Figure~\ref{figs:tf}, together with the ``best fit'' power-law advocated by
Giovanelli et al. (1997, data shown are compiled from that study as well as from
those of Mathewson, Ford \& Buchhorn (1992) and Han \& Mould 1992). The simulated galaxy
(solid circle) is seen to lie significantly below the mean relation, about $\sim
1$ magnitude fainter than most late-type spirals of similar rotation speed.

As discussed in \S~\ref{sssec:angmom}, a more telling description of the
discrepancy is to consider the offset as due to the increase in circular
velocity resulting from the presence of the highly-concentrated spheroidal
component. For example, if the spheroid was removed then the circular velocity
at $R=2.2\, R_{\rm d}$ would change from $V_{\rm rot}=260$ km s$^{-1}$ to $207$
km s$^{-1}$ (obtained by adding the
stellar disk, gas, and dark matter contributions, see Figure~\ref{figs:vcir1}b). 
This $\sim 20\%$ shift would be
enough to explain much of the velocity offset between the simulated galaxy and
late-type disks seen in Figure~\ref{figs:tf}. It also suggests that the presence
of a highly-concentrated spheroidal component in early type galaxies raise the
rotation speed of the luminous material, with the consequent systematic offset
in the Tully-Fisher relation.  Such interpretation is consistent with
Tully-Fisher data of spheroid-dominated S0 galaxies, which show a similar offset
relative to late-type disks (see filled triangles and diamonds in
Figure~\ref{figs:tf}), and shares the trend towards higher $V_{\rm rot}$/lower
$L$ reported by Kannappan, Fabricant \& Franx (2002) in early type spirals.

\section{Discussion}
\label{sec:disc}

The analysis described above provides direct confirmation that the origin of the
major photometric and dynamical components of spiral galaxies may be traced to
the mode and timing of their mass accretion history. Stars formed before the
merging activity of the galaxy is over assemble into a slowly-rotating,
centrally concentrated spheroid that extends out to several hundred kpc. Gas
accreting later settles into a thin, centrifugally supported disk of size
dictated by the available angular momentum and by the circular speed of the
system. Structurally, the stellar distribution in the simulated galaxy resembles
that of observed spirals, as demonstrated by the similarity in color and surface
brightness to Sab galaxy UGC615.

Overall, this should be regarded as a success for the $\Lambda$CDM scenario,
albeit a modest one, given that our simulation highlights as well a number of
concerns regarding the general viability of hierarchical galaxy formation
models. Two major (and related) ones refer to: (i) the prevalence of the dense
spheroidal component in the dynamics of the galaxy, and (ii) the dominance of
old stars in the stellar population of the system. Both concerns are rooted in
the rather lively merging history of the galaxy as well as in the high
efficiency of gas cooling within the earliest collapsing progenitors. These are
common features in the formation process of a galaxy assembled hierarchically,
and lead to a dominant spheroidal component even in a system (such as the one
considered here) chosen on the basis of its quiet accretion history at late
times. This underlines the concern that the frequency of spheroids and the age
distribution of stars may actually be inconsistent with the typical collapse and
merging history of galaxy-sized $\Lambda$CDM halos.

On a more optimistic note, our simulation does also provide convincing
evidence that these shortcomings might be resolved by adopting a
feedback recipe that is more efficient at preventing the
transformation of gas into stars in early collapsing
progenitors. Indeed, as discussed in \S~\ref{ssec:sclaws}, the disk
component on its own (disregarding the spheroid) has properties which
agree rather well with those of late-type spirals. Stripped of the
spheroid, the simulated galaxy would have a circular velocity speed of
$\sim 207$ km s$^{-1}$; a specific angular momentum of $j_{\rm
disk}=2\, R_d \, V_{\rm rot}=2.7 \times 10^3$ km s$^{-1}$ kpc; and a
total $I$-band magnitude of $-21.73$ (see Table~\ref{tab:gxphotdec}).

The angular momentum (size) and velocity of the disk would be in good agreement
with observation, as shown by the arrow in Figure~\ref{figs:vjdisk} which points
to the location of the galaxy after removal of the spheroid. The location of the
galaxy in the Tully-Fisher relation is less affected by this exercise (the
galaxy moves parallel to the observed relation, as shown by the arrow in
Figure~\ref{figs:tf}), perhaps signaling the need for genuinely lower stellar
mass-to-light ratios (see Eke, Navarro \& Steinmetz 2001 and
\S~\ref{ssec:dynprop}).  Finally, eliminating the spheroid would remove the
sharp central increase and subsequent decline in the rotation curve, improving
agreement with observation (see Figure~\ref{figs:vcir1}b).

Although this exercise is rather crude (it assumes, for example, that
the dark matter distribution would not be affected by the removal of
the spheroid, which is clearly not fully accurate) it still serves to
illustrate that the properties of the disk component, most of which is
assembled after $z \sim 1$, compares favorably with those of late-type
spirals. We conclude that full agreement with observation seems to
demand a mechanism that prevents the swift transformation of gas into
stars in early collapsing clumps and even perhaps allows for the
selective loss of this low angular momentum material, as argued by
NS00b.  Preliminary, but encouraging, results in this direction have
recently been reported by Sommer-Larsen, G\"otz and Portinari (2002)
with feedback schemes where the efficiency of the energetic input
increases with redshift. Progress in modeling galaxy formation in a
$\Lambda$CDM cosmogony thus appears to hinge on finding and
implementing a realistic and physically compelling description of star
formation and feedback.

\section{Summary}
\label{sec:conc}

We present a detailed analysis of the dynamical and photometric properties of a
disk galaxy simulated in the $\Lambda$CDM scenario.  The simulation is fully
3-dimensional and includes the gravitational and hydrodynamical effects of dark
matter, gas and stars. Star formation is modeled through a simple recipe that
transforms cold, dense gas into stars at a rate controlled by the local gas
density. Energetic feedback from stellar evolution is included, and calibrated
to match observed star formation rates in isolated disk galaxy models.

Our main results may be summarized as follows:

\begin{itemize}

\item{
The galaxy is assembled through a number of high-redshift mergers followed by a
rather quiescent period after $z \sim 1$. Our implementation of feedback is
rather inefficient at preventing stars from forming profusely in early
collapsing progenitors; $50\%$ of the present-day stars have already formed by
$z\sim 2.3$ and $75\%$ of them by $z\sim 1.3$.  }

\item{
At $z=0$, the $\sim 10^{11}\, M_{\odot}$ of stars in the galaxy are distributed
between two easily discernible components: a spheroid composed mainly of older
stars and a rotationally-supported disk of young stars. The surface brightness
profile can be fit adequately by the superposition of an $R^{1/4}$ spheroid and
an exponential disk; approximately half of the total light comes from each
component, although less than about a quarter of the stars form after the last
major merger.  This reflects the higher luminosity output of the younger stars
that make up the disk component.}

\item{
The stellar component has a mild age and color gradient, becoming bluer and
younger farther from the center: the luminosity-weighted age drops from $\sim 8$
Gyr at $R=1$ kpc to $\sim 6$ Gyr at $R=15$ kpc and its $B-R$ color drops from
$\sim 1.4$ to $\sim 1.0$ in the same radial range.  }

\item{
Photometrically, the simulated galaxy closely resembles the nearby Sab galaxy
UGC615 but their dynamical properties differ significantly. The simulated galaxy
rotates faster and has a declining rotation curve dominated by the spheroid near
the center. The decline in the circular velocity is at odds with observation and
results from the high concentration of the dark matter and stellar components,
as well as from the relatively high mass-to-light ratio of the stars.  }

\item{
The simulated galaxy lies $\sim 1$ mag off the I-band Tully-Fisher relation of
late-type disks, but is in reasonable agreement with Tully-Fisher data on S0
galaxies, suggesting that the systematic offset in the Tully-Fisher relation of
early-type spirals is due to the increase in circular speed caused by the
presence of the spheroidal component.}

\item{
The luminous component has a specific angular momentum well below that of
late-type spirals of similar rotation speed, in agreement with previous
simulation work. This reflects mainly the importance of the slowly-rotating
spheroidal component; the exponential scalelength of the disk ($R_d=5$ kpc) is
actually comparable to that of late-type spirals of similar rotation speed
($V_{\rm rot}\sim 260$ km s$^{-1}$). }

\end{itemize}

Most discrepancies with observation may be traced to the dominance of
the massive, dense, and slowly-rotating spheroidal component. On its
own, the disk component has properties rather similar to those of
late-type spirals: its exponential scalelength, colors, and luminosity
are all in reasonable agreement with galaxy disks of similar rotation
speed. Taken together, these results lend support to the view that
multi-component disk galaxies are formed naturally in hierarchically
clustering scenarios, where spheroids are the result of mergers and
disks of quiescent accretion.

Some worries remain, however, that fine tuning may be needed to reconcile model
predictions with observation. In particular, the early collapse and high merging
rates characteristic of scenarios such as $\Lambda$CDM might be difficult to
reconcile with the abundance of ``pure disk'' galaxies observed in the local
Universe, unless feedback is able to hinder star formation in high-redshift
progenitors much more efficiently than in the modeling we present
here. Accounting for the morphologies and abundance of galaxies in the
$\Lambda$CDM paradigm will likely require a better understanding of the way gas
cooling and accretion and star formation couple together, especially at high
redshift. Unraveling the puzzle of galaxy formation in a hierarchically
clustering universe remains a tantalizingly close, yet elusive, proposition.

\acknowledgments

We thank Sheila Kannappan for sharing and discussing unpublished
kinematic data for UGC615 (reproduced here courtesy of S.K. and D.
Fabricant).
This work has been supported by grants from the U.S. National Aeronautics and
Space Administration (NAG 5-10827), the David and Lucile Packard Foundation, the
Natural Sciences and Engineering Research Council of Canada, and Fundaci\'on
Antorchas from Argentina. 

\clearpage

\begin{deluxetable}{lccccccc}
\tablecaption{Photometric decomposition of simulated galaxy \label{tab:gxphotdec}}
\tablehead{
\colhead{Band} &
\colhead{[$L^{fit}_{tot}$]} &
\colhead{[$L_{\rm bulge}$]} &
\colhead{$R^{b}_{eff}$} &
\colhead{${\mu}_{eff}$} &
\colhead{$L_{\rm disk}$} &
\colhead{$R_{d}$} &
\colhead{$\mu^d_{0}$} 
\\ 
\colhead{} &
\colhead{[$10^{10}\, L_{\odot}$]} &
\colhead{[$10^{10}\, L_{\odot}$]} &
\colhead{[kpc]} &
\colhead{[$10^9\, L_{\odot}/$kpc$^2$]} &
\colhead{[$10^{10}\, L_{\odot}$]} &
\colhead{[kpc]} &
\colhead{[$10^7 \, L_{\odot}/$kpc$^2$]}
\\ 
}

\startdata
$U$ & $3.41$ & $1.77$ & $0.41$ & $4.71$ & $1.64$ & $5.51$ & $8.62$ \\
$B$ & $3.30$ & $1.70$ & $0.53$ & $2.67$ & $1.59$ & $5.25$ & $9.20$ \\
$V$ & $3.12$ & $1.62$ & $0.65$ & $1.69$ & $1.49$ & $4.95$ & $9.72$ \\
$R$ & $3.66$ & $1.93$ & $0.74$ & $1.57$ & $1.73$ & $4.77$ & $12.9$ \\
$I$ & $4.42$ & $2.38$ & $0.83$ & $1.54$ & $2.03$ & $4.64$ & $15.0$ \\
$K$ & $8.34$ & $4.78$ & $1.00$ & $2.10$ & $3.57$ & $4.41$ & $29.2$ \\
\enddata

\tablecomments{
Luminosities and radii assume face-on projection, $h=0.65$ and a Scalo Initial Mass Function. }
\end{deluxetable}

\begin{deluxetable}{rrccrcrcr}
\tablecaption{Properties of simulated galaxy at $z=0$\label{tab:gxprop}}
\tablehead{
\colhead{Radius} &
\colhead{$M_{tot}$} & 
\colhead{$M_{\rm dm}$} & 
\colhead{$M_{stars}$} &
\colhead{$M_{gas}$} &
\colhead{$V_{circ}$} & 
\colhead{$j_{\rm dm}$} & 
\colhead{$j_{stars}$} &
\colhead{$j_{gas}$} 
\\ 
\colhead{[kpc]} &
\colhead{} & 
\colhead{[$10^{10}\, M_{\odot}$]} & 
\colhead{} &
\colhead{} &
\colhead{[km s$^{-1}$]} & 
\colhead{} & 
\colhead{[km s$^{-1}$ kpc]} & 
\colhead{} 
\\ }
\startdata
$200$ & $85.5$ & $70.8$ & $12.5$ & $2.22$ & $136$ & $1065$ & $1624$ & $3644$ \\
$21$ & $26.19$ & $14.3$ & $10.3$ & $1.59$ & $232$ & $157.9$ & $439$ & $2680$ \\
\enddata
\tablecomments{All quantities are measured within the radii given in column [1]
and assume $h=0.65$. The virial radius is $r_{200}=207$ kpc. The mass per
particle is $m_{\rm dm}=1.86 \times 10^7 \, M_{\odot}$ for the dark matter. In
total, there are $38,054$ dark matter particles within $200$ kpc. Before any
star formation takes place the gas particle mass is $3.29 \times 10^6\,
M_{\odot}$. There are $127,815$ star particles within $200$ kpc, not all of the
same mass; the average star particle mass is $9.75 \times 10^5 \, M_{\odot}$.}
\end{deluxetable}

\begin{deluxetable}{lcccc}
\tablecaption{Main photometric properties of simulated galaxy \label{tab:gxphot}}
\tablehead{
\colhead{Band} &
\colhead{Luminosity} &
\colhead{Magnitude} &
\colhead{$R_{eff}$} &
\colhead{$R_{eff}$}  
\\ 
\colhead{} &
\colhead{[$10^{10}\, L_{\odot}$]} &
\colhead{} &
\colhead{[face-on]} &
\colhead{[edge-on]} 
\\ 
}

\startdata
$U$ & $2.54$ & $-20.40$ & $4.00$ & $2.20$ \\
$B$ & $2.76$ & $-20.62$ & $3.26$ & $1.93$ \\
$V$ & $2.89$ & $-21.32$ & $2.75$ & $1.74$ \\
$R$ & $3.60$ & $-21.97$ & $2.44$ & $1.60$ \\
$I$ & $4.57$ & $-22.57$ & $2.21$ & $1.48$ \\
$K$ & $9.36$ & $-24.15$ & $1.86$ & $1.27$ \\

\enddata

\tablecomments{Luminosities are measured within $21$ kpc. Magnitudes and radii
assume $h=0.65$ and a Scalo Initial Mass Function with upper and lower
mass cutoffs of $100\, M_{\odot}$ and $0.1\, M_{\odot}$, respectively. Radii are
projected half-light radii and are given in kpc. }
\end{deluxetable}

\end{document}